\documentclass[aps,twocolumn,superscriptaddress]{revtex4-2}

\usepackage{graphicx}
\usepackage{amssymb, amsmath,mathtools}
\usepackage{bbm}
\usepackage{enumerate}
\usepackage{hyperref}
\usepackage{dsfont}
\usepackage{color}
\usepackage{ulem}
\usepackage{xcolor}
\usepackage{dcolumn}% Align table columns on decimal point
\usepackage{wasysym} % for hexagon symbol
\usepackage{xspace}
\usepackage{ulem}

\newcommand{\ve}[1]{\boldsymbol{#1}}
\usepackage{orcidlink}

\graphicspath{{./}{./Figures/}}

\begin{document}
	
	\title{Phase transitions on the dark side of the Gross-Neveu model: Spontaneous $\textrm{O}(4N)$ symmetry breaking at repulsive coupling}
	
	\author{Gabriel Osiander Rein\, \orcidlink{0009-0008-0712-5814}}
	%\email{gabriel.rein@uni-wuerzburg.de}
	\affiliation{\mbox{Institut f\"ur Theoretische Physik und Astrophysik,
			Universit\"at W\"urzburg, 97074 W\"urzburg, Germany}}
	\affiliation{W\"urzburg-Dresden Cluster of Excellence ct.qmat, Am Hubland, 97074 W\"urzburg, Germany}
	%\altaffiliation[Also at ]{W\"urzburg-Dresden Cluster of Excellence ct.qmat, Universit\"at W\"urzburg, 97074 W\"urzburg, Germany}%Lines break automatically or can be forced with \\
	
	\author{Fakher F. Assaad\, \orcidlink{0000-0002-3302-9243}}
	\affiliation{\mbox{Institut f\"ur Theoretische Physik und Astrophysik,
			Universit\"at W\"urzburg, 97074 W\"urzburg, Germany}}
	\affiliation{W\"urzburg-Dresden Cluster of Excellence ct.qmat, Am Hubland, 97074 W\"urzburg, Germany}
	
	\author{Igor F. Herbut}%
	\affiliation{\mbox{Department of Physics,
			Simon Fraser University, Burnaby, British Columbia, Canada V5A 1S6}}
	
	%\homepage{http://www.}
	%\collaboration{W\"urzburg-Dresden Cluster of Excellence ct.qmat, Universit\"at W\"urzburg, 97074 W\"urzburg, Germany}%\noaffiliation

	\date{\today}% It is always \today, today,
	%  but any date may be explicitly specified
	
	\begin{abstract}
		
		Gross-Neveu model in 2+1 dimensions exhibits a continuous transition from gapless Dirac semimetal to the gapped quantum anomalous Hall (QAH) insulator at a finite (attractive) coupling, at which the inversion and time-reversal symmetry become spontaneously broken, and the flavor O($M$) symmetry remains preserved. A unification of leading order parameters of 2+1 dimensional $N$ four-component Dirac fermions collects all Lorentz-singlet mass-like fermion bilinears, except the one condensing in the QAH state, into an irreducible representation of the O($M=4N$), and predicts another phase transition in the Gross-Neveu model to occur at a strong (repulsive) coupling. Here, a fermionic auxiliary-field quantum Monte Carlo algorithm is employed in order to study a lattice realization of the Gross-Neveu field theory in the repulsive regime, where the sign problem is absent. We indeed find the O($4N$) symmetry breaking transition out of Dirac semimetal to occur and to be weakly first-order for $N=2$, relevant to graphene. The size of the discontinuity and the magnitude of the critical coupling, however, both grow with $N$. Adding a finite chemical potential is found to break the symmetry and cause superconductivity. These results are in broad agreement with the predictions of the unified field theory. Our lattice model also displays an interesting exact O($2N$) symmetry, a subgroup of the low-energy O($4N$), and has the ordered ground state with the order parameter that belongs to its $N(2N-1)$-dimensional representation. Other order parameters are also examined, and a certain hierarchy among those  that belong to different representations of the exact $\textrm{O}(2N)$ is observed.  
		
	\end{abstract}
	
	\maketitle
	
	\section{Introduction} 
	
	All two-dimensional electronic systems with massless relativistic (Dirac) spectrum at low energies exhibit an emergent enlarged orthogonal symmetry  \cite{HerbutMandal}. 
	Graphene, for example, has such a symmetry to be O($8$), which is reduced to U($4$) in presence of chemical potential or magnetic field.
	The mass-order-parameters \cite{HerbutQED1,HerbutQED2}, i.e.\ fermion bilinears whose finite expectation value gaps out the Dirac fermions while preserving Lorentz invariance, can be grouped into irreducible representations (irreps) of this larger symmetry.
	This way, all the mass-gaps except one  belong to the irreducible second-rank symmetric tensor, which for graphene, for example, is 35-dimensional \cite{Ryu, Herbut2012}. This maximal symmetry \cite{HuangLee} of a two-dimensional Dirac system is reduced by the electron-electron (Coulomb) interactions and the lattice, which ultimately dictate the nature of the broken symmetry state that emerges at strong coupling.  Specific quantum phase transitions into such broken-symmetry states have been a subject of numerous studies in the last two decades \cite{HerbutReview}. In particular, the semimetal-antiferromagnetic-insulator transition in the standard Hubbard model on the honeycomb lattice can be understood in terms of spontaneous breaking of rotational symmetry in presence of Dirac fermions \cite{Herbut2006, AssaadHerbut, HerbutReview, Sorella, Ma, Yang}.
	
	It is interesting to ponder the nature of the quantum phase transition and the corresponding broken symmetry in the maximally symmetric unified theory of {\it interacting} 2+1-dimensional Dirac fermions. Such a field theory turns out to be unique and to have been present all along: any relativistic field theory with $M$ complex fermions and the  concomitant O($M$) symmetry is equivalent to the celebrated original Gross-Neveu (GN) model  \cite{HerbutMandal}. The GN model, on the other hand, is known only to have the O($M$)-symmetry-preserving phase transition for strong interactions of certain sign, at which the inversion symmetry, which exists in addition to the O($M$), becomes spontaneously broken \cite{GrossNeveu, Zinn-Justin, Gracey, Erramilli}. It has been argued, nevertheless, that above certain threshold of the interaction of the opposite sign the O($M$) should  break spontaneously as well \cite{HerbutMandal}. The nature or even the existence of such a phase transition on what one could call a ``dark side" of the GN model is presently an open problem \cite{HanHerbut1, HanHerbut2, HanHerbut3}.
	
	In order to address the putative  O($M$)-symmetry-breaking phase transition we utilize the determinant Monte Carlo method to study a lattice model %
	defined on the honeycomb lattice---that naturally yields Dirac fermions in 2+1d relevant to graphene---whose construction and symmetries are detailed in Sec.~II. This lattice model has been
	argued to undergo the standard O($M$)-preserving GN phase transition below certain {\it negative} critical interaction and in the large-$M$  limit. We believe the model can be understood as a possible UV completion of the GN field theory. The attractive (negative) interaction in the lattice model, however, leads to the sign problem, and we therefore can efficiently simulate it only for {\it repulsive} (positive) interaction, which, incidently, corresponds to the previously unexplored interaction regime of the GN theory. 
	An additional and unusual feature of our model which seems interesting in its own right is that it has an exact O(2N) symmetry, even on the lattice. For $N=2$, for example, which corresponds to spin-1/2 fermions, the possible mass-like order-parameters fall into six-dimensional and nine-dimensional irreps, as well as into two further (reducible) ten-dimensional representations of the $\textrm{O}(4)$. The lattice $\textrm{O}(4)$ can be understood as the subgroup of the maximal $\textrm{O}(8)$ symmetry that emerges only in the low-energy limit. The ground state of the model turns out to belong to the irrep 6 \footnote{We denote a representation by its dimensions whenever there is no ambiguity, as customary.},  which consists of the three components of the Néel state, the charge density wave, and two components of the s-wave superconductor. This should be contrasted with the Hubbard model, for example, where the interaction term already reduces $\textrm{O}(4)$ to $\textrm{SO}(4)$, and the same irrep 6 decomposes into two three-dimensional irreps \cite{YangZhang}.
	
	The paper is organized as follows: in the next section we define the lattice model and show it possesses an $\textrm{O}(2N)$ symmetry. In sec. III we discuss the physics of the model for attractive interaction, and argue that it has the GN transition into quantum anomalous Hall (QAH) state at a finite coupling. In sec. IV other mass-terms relevant for repulsive interactions are introduced. The results of the sign-problem-free Monte Carlo calculation for repulsive interaction are presented in sec. V. Further discussion is given in sec. VI. Some technical details are relegated to the Appendix. 
	
	\section{Lattice model} 
	
	We define the Hamiltonian as a sum of the standard hopping term and the new interaction term:
	\begin{equation}
		H = 2 t \sum_{\langle \ve{i},\ve{j} \rangle} c_{\ve{i}}^\dagger  c_{\ve{j}} + \frac{2 \lambda}{N} \sum_{\varhexagon} \{ \sum_{\langle \langle \ve{i},\ve{j} \in \varhexagon \rangle \rangle } \nu_{\ve{i}\ve{j} } c^\dagger _{\ve{i} } c_{\ve{j} } \} ^2, 
		\label{eq:Model}
	\end{equation}
where $c_{\ve{i}} ^\dagger = (c^\dagger _{\ve{i}, 1}, ... ,c^\dagger _{\ve{i}, N})$ creates a $N$-component complex fermion at site $\ve{i}$ on the honeycomb lattice. The phase factors $\nu_{\ve{i}\ve{j}} = -\nu _{\ve{j}\ve{i}} =\pm i$, with the signs as given in the Haldane model \cite{Haldane}. Both $t$ and $\lambda$ are real parameters, and $H$ is Hermitian. Kinetic energy describes the usual nearest-neighbor hopping from one triangular sublattice of the honeycomb lattice to the other, whereas the phase factors $\nu_{\ve{i}\ve{j}}$ in the interaction term connect the next-nearest-neighbor sites on the same sublattice, called hereafter A and B. The model is invariant under the global transformation $c_{\ve{i}}  \rightarrow U c_{\ve{i}} $, with $U \in \textrm{U}(N)$.
	
	It is convinient to introduce Majorana lattice fermions at each site $\ve{i}$, $\phi_{k,\ve{i}}$, $k=1,2$ as
	\begin{equation}
		c_{\ve{i} } = \frac{(-i)^n}{2} ( \phi _{1, \ve{i}} + i \phi_{2,\ve{i}}),
	\end{equation}
	\begin{equation}
		c_{\ve{i}} ^\dagger = \frac{i ^n}{2} ( \phi _{1, \ve{i}} ^T - i \phi_{2,\ve{i}} ^T ),
	\end{equation}
	with $n=0$ on the sublattice A, and $n=1$ on the sublattice B. Due to the phase factors $\nu_{\ve{i}\ve{j}}$ being imaginary {\it both} the kinetic and the interaction terms in the Hamiltonian now adopt simpler forms, 
	\begin{equation}
		H= i t  \sum_{\langle \ve{i},\ve{j} \rangle} \phi _{\ve{i}} ^T  \phi_{\ve{j}} + \frac{\lambda}{2 N } \sum_{\varhexagon} \{ \sum_{\langle \langle \ve{i},\ve{j} \in \varhexagon \rangle \rangle } \nu_{\ve{i}\ve{j} } \phi ^T _{\ve{i}} \phi_{\ve{j}}\} ^2,
	\end{equation}  
	where $\phi_{\ve{i}} ^T = (\phi_{1,\ve{i},1}, \phi_{2,\ve{i},1}... \phi_{1,\ve{i},N}, \phi_{2,\ve{i},N})^T  $ is a $2 N$-component (``real") Majorana fermion. The lattice Hamiltonian $H$ is now manifestly symmetric under the (global) transformation
	\begin{equation}
		\phi_{\ve{i} }\rightarrow O \phi_{\ve{i}} , 
	\end{equation}
	with $O \in \textrm{O}(2N)$.  
	
	For $N=2$ (spin-1/2) the generators of the $\textrm{SO}(4)= \textrm{SU}_s (2) \times \textrm{SU}_c (2) \subset \textrm{O}(4) $ are $S_a = (  \sigma_3 \otimes \sigma_2, \sigma_1 \otimes \sigma_2, \sigma_2 \otimes 1_2)/2$  and  $C_a = ( \sigma_2 \otimes \sigma_3, \sigma_2 \otimes \sigma_1,  1_2 \otimes \sigma_2)/2$, $a=1,2,3$. $\sigma_i$ are the usual Pauli matrices, and $1_2$ is the two-dimensional unit matrix. The first set generates spin-rotations $\textrm{SU}_s (2)$, and the particle-number $C_3$ generates the
	gauge $\textrm{U}(1)\subset \textrm{SU}_c (2)$ 
	(corresponding to charge conservation)
	, for example. 
	
	The model shares the same symmetry with the recent proposal of Ref. \cite{Kaul}, but differs in the interaction term. 
	We note for later reference that the
    integer $M$ used in the introduction is related to the lattice parameter $N$ by $M=4N$ ($N$ spin times two Majorana species times two valley degrees of freedom).

	\section{QAH state and the GN transition} 
	\label{sec:QAH}
	The goal of this section is to establish, by a saddle-point analysis and a low-energy expansion, that the lattice model \eqref{eq:Model} undergoes the standard O($M$)-preserving Gross-Neveu transition into the QAH insulator for sufficiently strong attractive coupling $\lambda<0$. In contrast, for $\lambda>0$ the saddle-point approximation does not predict a transition.

	After the Hubbard-Stratonovich transformation \cite{Negele} of the interaction term the (finite-temperature) action that corresponds to the above Hamiltonian, written in Fourier space, becomes
	\begin{widetext}
		\begin{eqnarray}
			S = \frac{N T}{2|\lambda|} \sum_{\ve{q}, \omega} \chi(\ve{q}, \omega) \chi(-\ve{q}, -\omega) + 
			T \sum_{\ve{k},\omega} \phi ^T (-\ve{k}, -\omega) [ 1_{2N} \otimes ( i\omega + t (f_1 (\ve{k}) \sigma_2 + f_2 (\ve{k}) \sigma_1 ) ) ] \phi ( \ve{k},\omega) +  \\ \nonumber 
			i^m T^2 \sum _{ \ve{k},\omega, \ve{q}, \nu}  g(\ve{k}) \chi( \ve{q},\nu) \phi ^T (-\ve{k} -\ve{q}, -\omega-\nu)  (1_{2N} \otimes \sigma_3)  \phi(\ve{k},\omega),  
		\label{Eq.Hubbard-Stratonovich}
		\end{eqnarray}
	\end{widetext}
	with $f_1 (\ve{k}) + i f_2 (\ve{k}) = \sum _{\ve{b}_i} \exp (i \ve{k}\cdot \ve{b}_i)$ and $g(\ve{k}) = 2 \sum_{\ve{a}_i} \sin ( \ve{k}\cdot \ve{a}_i)$, where $\ve{b}_{1,2,3}$ are the vectors that connect a site to its nearest neighbors, and $\ve{a}_3= \ve{b}_2 - \ve{b}_1$ +  cyclic permutations, connect the next-nearest-neighbors. The exponent $m$ in the last term is $m=0$ if $\lambda<0$ (attraction), and $m=1$ if $\lambda >0$ (repulsion). Since the honeycomb's unit cell consists of two elements, the Majorana fermion $\phi(\ve{k},\omega)= (\phi_A (\ve{k},\omega), \phi_B (\ve{k},\omega) )^T $, and now has twice as many components as before, which is $4 N$. The sums over the momenta $\ve{k}$ and $\ve{q}$ are taken over the first Brillouin zone, and $\omega$ are the usual fermionic Matsubara frequencies.
%\fa{In Equation 4  $\phi$  has 2N component,  then in the above equation it has 4N components.  Could we use $\gamma$ on the lattice. No strong feelings about this but I found it confusing.}
	
	For attractive interaction $\lambda<0$ and $N\rightarrow \infty$ the saddle-point contribution to the partition function yields the exact result, and the Hubbard-Stratonovich field is determined by the gap equation
	\begin{equation}
		\chi = -4 \lambda T \sum_{\ve{k},\omega} \frac{\chi g(\ve{k})^2 }{\omega^2 + t^2 (f_1 (\ve{k}) ^2 + f_2 (\ve{k})^2) + \chi^2 g(\ve{k})^2}
	\end{equation}
	where $\chi (\ve{k},\omega) = (\chi/T) \delta_{\ve{k},0} \delta_{\omega,0} $.
	The integral is dominated by contributions from the vicinity of the two Dirac points at  $\ve{k}=\pm \ve{K}$, and has a non-trivial solution only for  $\lambda < \lambda_c$, with $\lambda_c<0$. For $\lambda >\lambda_c$, and in particular for all $\lambda >0$, the solution of 
	the gap equation  
	is only the trivial $\chi=0$. 
	
	To go  beyond  $N\rightarrow \infty$ limit we integrate the fluctuations of the uniform Hubbard-Stratonovich field, and zoom in on the low-energy modes near the two Dirac points. 
	The energy-momentum dispersion on the honeycomb lattice is known to be linear over a good portion of the honeycomb lattice \cite{HerbutReview}.
	At low energies this way one finds the Lagrangian 
	\begin{widetext}
		\begin{eqnarray}
			L = \psi_L ^T (\partial_\tau + ( \sigma_3 \otimes 1_{2N} \otimes \sigma_2)  i \partial_1 + ( 1_2 \otimes 1_{2N} \otimes \sigma_1) i \partial_2) \psi_R
			+ \frac{ \lambda g(\ve{K}) ^2 }{2N} ( \psi_L ^T \sigma_3 \otimes 1_{2N} \otimes \sigma_3 \psi_R )^2  + O ( \psi_L ^T \partial ^2 \psi_R), \label{eq:L}
		\end{eqnarray}
	\end{widetext}
	where 
	\begin{equation}
		\psi_R (\ve{x},\tau) = T \sum_{|\ve{q}| \ll \Lambda, \omega} e^ {i( \ve{q}\cdot \ve{x} +  \omega \tau)} (\phi (\ve{K}+\ve{q}, \omega) , \phi (-\ve{K} +\ve{q}, \omega) )^T,  \label{eq:psi_R}
	\end{equation}
	and 
	\begin{equation}
		\psi_L ^T  (\ve{x},\tau) = T \sum_{|\ve{q}| \ll \Lambda, \omega} e ^ { i ( \ve{q}\cdot \ve{x} +  \omega \tau)} (\phi (-\ve{K}+\ve{q}, \omega) , \phi (\ve{K} + \ve{q}, \omega) ), \label{eq:psi_L}
	\end{equation} 
	are the Majorana $8N$-component fields which collect together the low-energy fermionic modes. The final doubling of the number of degrees of freedom comes from having two Dirac points. $\Lambda$ is the UV cutoff, arbitrary, but assumed much smaller than the inverse of the lattice spacing. 

Apart from the irrelevant terms, the low-energy theory has the structure of the GN model. The transition below the (negative) critical coupling $\lambda_c$ should be into the ground state with 
	\begin{equation}
		\langle \psi_L ^T (\sigma_3 \otimes 1_{2N} \otimes \sigma_3 ) \psi_R \rangle \neq 0, \label{eq:QAH-mass-term}
	\end{equation}
	at which fermions would become gapped and the time reversal and the sublattice symmetries broken. The exact O($2N$) symmetry of the lattice model, and the emergent O($4N$) symmetry of the low-energy field theory to be extracted shortly,  in contrast, at this transition remain preserved. The symmetry breaking pattern and the field content therefore suggest that the semimetal-insulator transition at strong attractive interactions should be described by the GN continuum field theory, and may be expected therefore to be continuous at finite $N$. The broken-symmetry phase can be recognized as the QAH insulator \cite{Haldane}.  
	
	In summary, in the attractive regime $\lambda<0$ the model flows to the GN fixed point and undergoes an O($4N$)-preserving transition into the QAH state. The sign problem prevents direct Monte Carlo access to this regime, which motivates the study of the repulsive side in the following sections. The saddle-point approximation shows no solution on this side but the next section explains that this is not the end of the story.
	\section{Masses on the dark side} 
	Motivated by a Fierz rearrangement described in Ref. \cite{HerbutMandal}, the goal of this section is to classify all Lorentz-invariant, mass-like order parameters---other than the QAH of Sec.~\ref{sec:QAH}---and to combine them into an irrep of the SO($4N$) symmetry group. A spontaneous breaking of this symmetry has been speculated to cause a transition on the dark side of the GN model.
	
	At an attractive interaction $\lambda<0$ the lattice model suffers from the sign problem, which prevents us to efficiently study what appears to be the usual criticality in the GN field theory by Monte Carlo. At the same time this implies that there is no sign problem when $\lambda >0$ \cite{WuZhang}. For positive $\lambda$, and at least at large $N$, 
	based on the gap-equation 
	one expects that the QAH order parameter $\chi =0$. 
	It has recently been argued, however, that the GN field theory suffers a different phase transition above a certain repulsive (positive) critical $\lambda$, at which the internal symmetry O($4N$), hidden in the above low-energy Lagrangian, spontaneously breaks down to $\textrm{O}(2N) \times \textrm{O}(2N)$ \cite{HerbutMandal, HanHerbut1}. This transition which should occur at strong repulsive interactions in our model is the central theme of this work.
	
	%\fa{Would it not be a good idea to repeat the Fierz identity  you  refer to in the conclusion.  I found this equation very useful. In particular, I would have in mind Eq. 15 of  \href{https://arxiv.org/pdf/2406.01681}{this}  paper.  Up to you Igor if you want to include it or not.  }

	To discern the mass-like order-parameters other than QAH, which are relevant to the repulsive regime, we begin by examining the fermion bilinear that defines the QAH order parameter more closely. It has the form  $ \sim \psi_L ^T M \psi_R $, with the particular matrix $M= \sigma_3 \otimes 1_{2N} \otimes \sigma_3 $. Since by our definition in Eqs. (\ref{eq:psi_R}) and (\ref{eq:psi_L}) $\psi_R = (\sigma_1 \otimes 1_{2N}\otimes 1_2) \psi_L$, the matrix $M$ satisfies the constraint: 
	\begin{equation}
		(\sigma_1 \otimes 1_{2N}\otimes 1_2) M ^T (\sigma_1 \otimes 1_{2N}\otimes 1_2 ) = - M,  
	\end{equation}
which replaces the usual condition of antisymmetry of $M$ when it stands in between identical Majorana fields on the left and the right.
If we demand that: a) $M$ is a Hermitian mass-matrix that anticommutes with the (linearized) Dirac Hamiltonian, and b) that it describes a translationally uniform state, i. e. that it also commutes with $\sigma_3 \otimes 1_{2N}\otimes 1_2$ \footnote{The
		generator $\sigma_3\otimes\mathbf{1}_{2N}\otimes\mathbf{1}_2$ acts on the valley degree of freedom and distinguishes the two Dirac points at $\pm K$. A translationally uniform mass matrix must not mix modes at $+K$ and $-K$, which is equivalent to requiring that $M$ commutes with this generator.}
\cite{HerbutJuricicRoy}, it becomes either 
	\begin{equation}
		M= \sigma_3 \otimes S_{2N} \otimes \sigma_3, 
	\end{equation}
	or 
	\begin{equation}
		M= 1_2 \otimes A_{2N} \otimes \sigma_3, \label{eq:M2}
	\end{equation}
where $S_{2N}$ ($A_{2N}$) is a real, symmetric (imaginary, antisymmetric) $2N \times 2N$ matrix. $A_{2N}$ ($S_{2N}$) transforms as the two-component antisymmetric (symmetric) tensor under the exact O($2N$). Imposing Tr$S_{2N} =0$ makes the representation irreducible by eliminating the unit matrix $1_{2N}$, which as a scalar under O($2N$) yields the QAH state in Eq. (\ref{eq:QAH-mass-term}).  
	
It may be instructive to consider the smallest case of $N=2$ again. The antisymmetric representation is six-dimensional, and given by the set of generators $S_a$ and $C_a$ already introduced in sec. II. We therefore find that the matrices $C_1$ and $C_2$ form a two-component vector under the action of the 
	particle-number-generator $C_3$, but also that they are both invariant under $\textrm{SU}_s (2)$. They represent therefore the $x$- and $y$- components 
	(reflecting the real and imaginary part rotating as a vector under U(1))
	of the s-wave superconductor.
	$C_3$, on the other hand, being also invariant under $SU_s(2)$ and of course commuting with itself, yields the charge density wave, i. e. the density imbalance between the A and B sublattices. Similarly, the generators of $SU_s (2)$ correspond to the three components of the Néel order parameter. Deconstructing $\psi_L ^T M \psi_R $ with $M$ in the form as in Eq. (\ref{eq:M2}) confirms these assignments and also reveals these fermion bilinears to be local, on-site, lattice terms. 
	
	The irreducible symmetric representation for $N=2$ is nine-dimensional; anticipating the correct identifications, let us define $E = (\sigma_1 \otimes 1_2, \sigma_3 \otimes 1_2, \sigma_2 \otimes \sigma_2)$, $F_{x}= (\sigma_3 \otimes \sigma_3, \sigma_1 \otimes \sigma_3, 1_2 \otimes \sigma_1)$, and $F_{y}= (\sigma_3 \otimes \sigma_1, \sigma_1 \otimes \sigma_1, 1_2 \otimes \sigma_3)$. All three, $E$, $F_x$, and $F_y$ are then three-component vectors under spin-rotations. $E$ is an (particle-number-preserving) insulator, and $F_x$ and $F_y$ are components of a (particle-number-violating) superconductor. Writing the nine bilinears explicitly identifies $E$ as the quantum spin Hall (QSH) order parameter \cite{KaneMele}, and $F$ as the six components of the spin-triplet ``f-wave" superconductor \cite{Honerkamp}. All nine fermion bilinears translate into next-nearest-neighbor lattice terms. 
	
	Relaxing the uniformity condition on the mass-matrices allows two more possibilities: 
	\begin{equation}
		M= \sigma_k \otimes S_{2N} \otimes \sigma_2, 
	\end{equation}
	$k=1,2$, which transform into each other under the action of the generator $\sigma_3 \otimes 1_{2N}\otimes 1_2$. These additional mass-like order parameters represent Kekulé bond-density-waves, both insulating \cite{Hou} and superconducting \cite{Roy}. For $N=2$ there are evidently 2 $\times$ 10 = 20 of these mass-matrices. Altogether, therefore, for $N=2$ there are 1+6+9+20=36 possible mass terms, including the QAH state  \cite{Ryu, Herbut2012}. Kekulé orders belong to a representation which is actually reducible under $\textrm{O}(2N)$, and to make it irreducible again the condition $Tr S_{2N}$ would need to be imposed. We need not make this distinction, however, since all the Kekulé orders will turn out to always stay short-ranged. 
	
	\begin{figure*}
		\centering
		\includegraphics[width=0.95\textwidth]{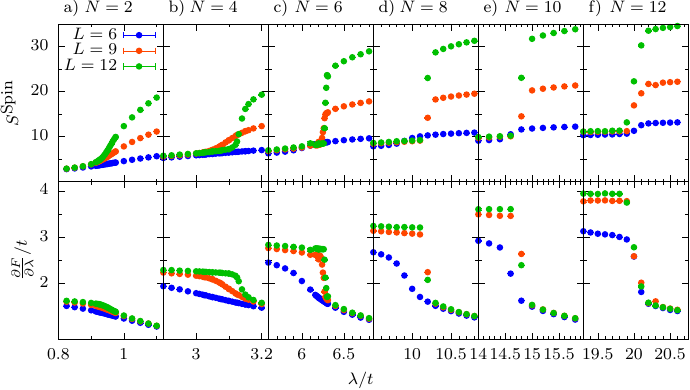}
		\caption{Spin structure factor at $\boldsymbol{q}=0$ (top) and scaled potential energy (bottom) as a function of $\lambda/t$ for different values of $N$. The structure factor indicates a transition from a Dirac SM to an AFM insulator that gets shifted towards larger $\lambda$ with increasing $N$. The potential energy shows a discontinuity, which becomes more pronounced as $N$ increases. In (c), (d), (e) and (f) the scale of the $x$-axis is adjusted in order to better resolve the split of the data with different lattice sizes. 
		If no error bars are visible, the statistical errors are smaller than the symbol sizes throughout.}
		\label{fig:structurefactor}
	\end{figure*}

	By finding all non-trivial commutators between pairs of the identified mass-matrices one may obtain the generators of the group of symmetry of the Dirac (linearized) interacting theory in Eq. (\ref{eq:L}):
	\begin{eqnarray}
		\sigma_3 \otimes S_{2N} \otimes 1_2, \\
		1_2 \otimes A_{2N} \otimes 1_2, \\
		\sigma_k \otimes A_{2N} \otimes \sigma_1. 
	\end{eqnarray}
	$k=1,2$. For $N=2$ there are therefore 10 + (3 x 6) = 28 generators, for example. For general $N$ there are $2N (4N-1)$ generators, and the resulting Lie algebra is SO($4N$).  For any $N$ the generators commute with the QAH mass-matrix, and therefore the representation is reducible. One can show it is an orthogonal sum of two equivalent irreps of the dimension $4N$. For $N=2$ we thus arrive at the spinor representation of the  SO($8$), previously derived directly from the Dirac Hamiltonian  \cite{HerbutMandal}. 
	It is straightforward to check that the algebra of generators of SO($4N$) is closed, as well as that the commutators of the generators with the mass-matrices produce only other mass-matrices. This confirms that the masses indeed provide a representation of the extracted Lie algebra.
	
	In summary, the mass-like order parameters of the low-energy theory yield an irrep of SO($4N$), and the candidates that lead to the transition on the dark side are the six-dimensional and nine-dimensional irreps	of the exact lattice O($2N$), plus the (reducible) Kekul\'{e} representations. The Monte Carlo study of the resulting phase transition is subject to the next section.
	
	\section{Results \label{sec:Results}}
	\begin{figure*}
		\centering
		\includegraphics[width=0.9\textwidth]{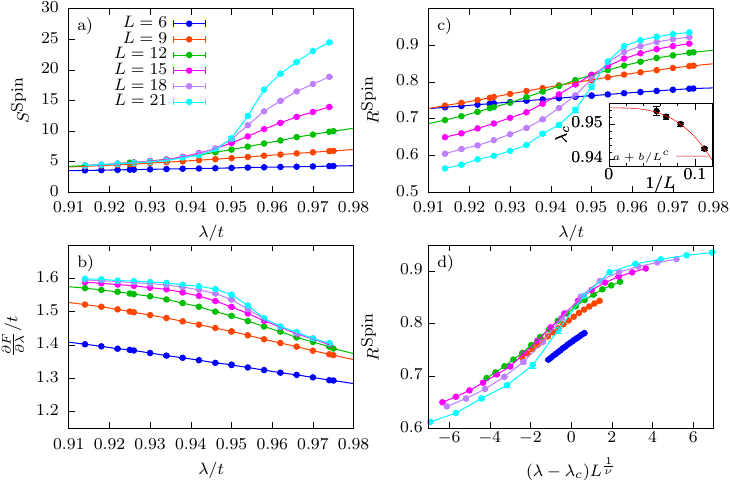}
		\caption{Spin-structure factor at $\boldsymbol{q}=0$ (a), derivative of free energy with respect to $\lambda$ (b) and correlation value $R$ at $\boldsymbol{q}=0$ (c) as a function of interaction strength $\lambda/t$ for $N=2$ at $\beta t=L$. The inset in (c) shows the crossing point $\lambda_c$ between the correlation ratio for a respective lattice size $L$ and $L+3$ as a function of $1/L$ together with a power law fit. Thereby, in the thermodynamic limit we extract a crossing point $\lambda_c=0.955(3)$. With this and a correlation length exponent $\nu=0.529$ we obtain our best data collapse shown in (d).
			}
		\label{fig:transition_N_2}
	\end{figure*}
	We now proceed to study the model defined in equation \eqref{eq:Model} by employing a finite temperature  auxiliary-field Quantum Monte Carlo (QMC) algorithm \cite{QMC-PhysRevD.24.2278,QMC-PhysRevB.40.506,Assaad2008-QMC} based on the ALF package \cite{ALF_v2.4}. 
	In this approach, since the interaction term is written as a perfect square it can be  decoupled via a Hubbard-Stratonovich transformation with space ($\ve{i}$) and time ($\tau$) dependent auxiliary field $\chi (\ve{i},\tau)$ as in Eq.~\ref{Eq.Hubbard-Stratonovich}.  With this, and after integrating out the fermionic degrees of freedom the partition function reads
	\begin{equation}
		\label{Eq:QMC}
		Z =  \int D \left\{ \chi (\ve{i},\tau) \right\} e^{- N S\left\{ \chi (\ve{i},\tau) \right\}}.
	\end{equation}
	Due to  requisite time reversal symmetry the action is real for  $\lambda>0$ \cite{WuZhang}, so that  $e^{- N S\left\{ \chi (\ve{i},\tau) \right\}}$ is a viable  weight function for  Monte Carlo importance sampling.
	We set $t=1$ for all of the presented calculations and, if not stated otherwise, consider a half-filled band, i. e. one electron per site, which corresponds to vanishing chemical potential. Furthermore, we use a symmetric Trotter decomposition with an imaginary time discretization $\Delta\tau=0.4/N$ and employ a finite temperature code at $\beta=L$ unless otherwise specified. 
	These choices are elaborated on in the Ref.\cite{Rein} which uses an implementation that closely resembles the case studied here. The QMC method is described in more detail in the appendix.
	In order to resolve a possible transition in the lattice model for $\lambda>0$, we calculate the structure factor
	\begin{align}
		S_{\delta,\delta'}^{O}(\boldsymbol{q})=\frac{1}{L^2}\sum_{\boldsymbol{r},\boldsymbol{r}'}&e^{i\boldsymbol{q}\cdot(\boldsymbol{r}-\boldsymbol{r}')}\big(\big\langle\hat{\boldsymbol{O}}_{\boldsymbol{r},\delta}\hat{\boldsymbol{O}}_{\boldsymbol{r}',\delta'}\big\rangle \nonumber\\
		&- \big\langle\hat{\boldsymbol{O}}_{\boldsymbol{r},\delta}\big\rangle\big\langle\hat{\boldsymbol{O}}_{\boldsymbol{r}',\delta'}\big\rangle\big), 
		\label{eq:structurefactor}
	\end{align}
	where $L^2$ is the lattice size and $O$ is an observable defined by the order parameter $\hat{\boldsymbol{O}}_{\boldsymbol{r},\delta}$, with $\boldsymbol{r}$ and $\delta$ being the position of a unit cell and the number of an orbital respectively. If not stated otherwise we calculate the trace of the structure factor with respect to the orbital $\delta$. In Fig. \ref{fig:structurefactor} we capture the antiferromagnetic (AFM) insulating state by choosing the vector order parameter $\hat{\boldsymbol{O}}=\boldsymbol{c}^\dagger \boldsymbol{T}\boldsymbol{c}$ in eq. (\ref{eq:structurefactor}) where $\boldsymbol{T}$ is an $N^2-1$-component vector containing the generators of SU($N$). The diverging behavior of the structure factor with increasing lattice size signals the O($2N$) symmetry broken phase. It can be observed that for $N> 4$ the structure factor, at least for $L=9$ and $L=12$, exhibits a sharp jump at the transition point $\lambda_c$, which itself shifts to larger values with the increase of $N$. The jump becomes better defined and more sensitive to the lattice size as $N$ increases. In the bottom row of Fig. \ref{fig:structurefactor} the potential energy $E_{\textrm{pot}}=\langle H_{\lambda}/\lambda\rangle$ is plotted. Since at $T=0$ the internal energy of the system can be identified with the free energy, the potential energy equals the derivative of the free energy with respect to $\lambda$. Thus, a jump in the potential energy also serves as an indicator of a first-order transition, as the free energy typically then shows a kink. 
	
	The $N=2$ case which corresponds to spin-1/2 fermions requires a closer look. For the lattice sizes shown in Fig. \ref{fig:structurefactor} the discontinuity at the transition is not clearly discernable. Nevertheless, simulating larger lattices suggests that the transition may even here still be (weakly) first order. In Fig. \ref{fig:transition_N_2}, the structure factor and the free energy derivative are shown for systems up to $L=21$. The correlation ratio $R=1-\frac{S(\boldsymbol{Q}+\Delta\boldsymbol{q})}{S(\boldsymbol{Q})}$ with ordering wave vector $\boldsymbol{Q}$ and $|\Delta\boldsymbol{q}| = \frac{4\pi}{\sqrt{3L}}$ is plotted as a function of $\lambda$ in Fig. \ref{fig:transition_N_2}(c). The correlation ratio allows us to extract the crossing point $\lambda_c$ by scaling the crossing of the data for $L$ and $L+3$ to infinite system sizes. This can be observed in the inset of Fig. \ref{fig:transition_N_2}(c) together with a fit of the data by the function $a+b/L^c$. Hereby, we obtain a crossing point $\lambda_c=0.955(3)$. 
	The best data collapse plotted in \ref{fig:transition_N_2}(d) is then found for the correlation length exponent $\nu=0.529$. This is an unusually low value, and suspiciously close to the likely lower bound of 1/2 \cite{Vicari}. We note that it satisfies, but it is even closer, to the stricter bound of $\nu > 0.511$ for a unitary conformal field theory in 2+1D with only one tuning parameter. \cite{Nakayama} 
	The free energy derivative in Fig. \ref{fig:transition_N_2}(b) shows a rather abrupt change for $L=21$ and it becomes apparent that, close to the transition, with increasing lattice size it rises faster outside ($\lambda=0.95$) than inside the ordered phase ($\lambda=0.96$). This, together with a observed poor data collapse suggests a discontinuous transition probably taking place already at $N=2$.

	A finite chemical potential $\mu$ breaks the O($2N$) symmetry in favor of the superconducting order parameters,\cite{HerbutMandal} 
and the ordered ground state at its finite value is expected to be a superconductor.  The corresponding order parameter may be written as $\hat{O}_{\boldsymbol{i}}^{\textrm{SC}}=\frac{1}{2}c_{\boldsymbol{i},\sigma}^\dagger Y_{\sigma,\sigma'} c_{\boldsymbol{i},\sigma'}^\dagger+\textrm{h.c.}$, with $Y$ as an antisymmetric complex matrix. One does not need to be more specific here,  since the structure factor is independent on the precise form of $Y$. We provide a more detailed discussion in the appendix. In Fig. \ref{fig:doping} the particle number $\langle \hat{N}\rangle$ as well as the spin and SC structure factors are shown as a function of $\mu$. Note, that at $\mu=0$ the structure factors of spin and SC order parameters are identical, since they belong to the same irrep of the exact O($2N$) symmetry of the lattice Hamiltonian. For a small but finite chemical potential the O($2N$)-symmetry-broken phase quickly flops into a pure SC. For this calculation, a $\beta t=L^2$ scaling is employed assuming a dynamical critical exponent $z=2$ that reflects the quadratic shape of the gapped energy dispersion. At small chemical potential $\mu$ finite size effects may distort the data as can be observed in Fig. \ref{fig:doping}(a,b), especially for the small system sizes. 
	
	\begin{figure}
		\centering
		\includegraphics[width=0.9\columnwidth]{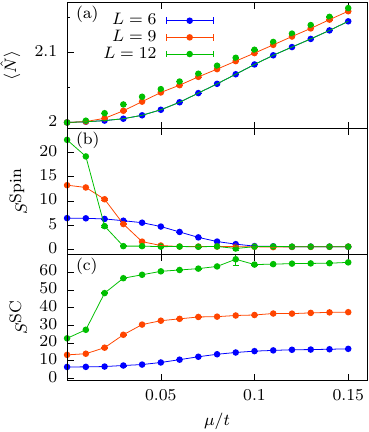}
		\caption{Particle number (a), Spin structure factor (b) and SC structure factor (c) at $\boldsymbol{q}=0$ as a function of chemical potential $\mu/t$ for $N=2$ at $\lambda=0.4t$ and $\beta t=L^2$.}
		\label{fig:doping}
	\end{figure}
	
Finally, in order to compare different possible orders in the system we compute the normalized structure factor
	\begin{equation}
		S_N^O(\boldsymbol{k}) = \frac{S^O(\boldsymbol{k})}{\sum_{\boldsymbol{k}\in\textrm{B.Z}} S^O(\boldsymbol{k})}, 
	\end{equation}
	and use it as a measure of the pronouncement of the peak of the structure factor relative to the background. In Fig. \ref{fig:S_k} the normalized structure factor is plotted for various momenta against the interaction strength $\lambda/t$ around the transition at $N=2$. In Fig. \ref{fig:S_k}(a) we see that at the ordering wave vector of the s-wave SC order parameter, namely $\boldsymbol{k}=0$, the structure factor increases rapidly relative to the overall Brillouin zone as the system undergoes the phase transition. This is not the case for the other momenta along the $k_x$-axis, reflecting a strong peak of the structure factor at $\boldsymbol{k}=0$. In accord with Fig. \ref{fig:structurefactor}(a) and Fig. \ref{fig:transition_N_2} this marks the transition to the O(4)-symmetry broken phase. Moreover, we plot the normalized structure factors of the order parameters of QSH $\hat{\boldsymbol{O}}_{\boldsymbol{r},\delta}^{\textrm{QSH}}=\nu_{\boldsymbol{i}_{\delta}\boldsymbol{j}_{\delta}} \boldsymbol{c}_{\boldsymbol{r}+\boldsymbol{i}_{\delta}}^\dagger \boldsymbol{\sigma} \boldsymbol{c}_{\boldsymbol{r}+\boldsymbol{j}_{\delta}}+\textrm{h.c.}$ and f-wave SC $\hat{\boldsymbol{O}}_{\boldsymbol{r},\delta}^{\textrm{fSC}}=\nu_{\boldsymbol{i}_{\delta}\boldsymbol{j}_{\delta}}' \boldsymbol{c}_{\boldsymbol{r}+\boldsymbol{i}_{\delta}}^\dagger \boldsymbol{\sigma}\sigma_2 \boldsymbol{c}_{\boldsymbol{r}+\boldsymbol{j}_{\delta}}^\dagger+\textrm{h.c.}$ where $\nu_{\boldsymbol{i}\boldsymbol{j}}'$ relates to $\nu_{\boldsymbol{i}\boldsymbol{j}}$ by a negative sign on sublattice B and $\delta$ specifies the six bonds of a honeycomb with legs $\ve{i}_{\delta}$ and $\ve{j}_{\delta}$. Singlet Kekulé $\hat{O}_{\boldsymbol{r},\delta}^{\textrm{sK}}=\boldsymbol{c}_{\boldsymbol{r}}^\dagger\boldsymbol{c}_{\boldsymbol{r+\boldsymbol{\delta}}}$ (here $\ve{\delta}$ is the relative position of the three nearest neighbors) as well as the $z$-component of triplet Kekulé $\hat{O}_{\boldsymbol{r},\delta}^{\textrm{tK}}=\boldsymbol{c}_{\boldsymbol{r}}^\dagger\sigma_3\boldsymbol{c}_{\boldsymbol{r+\boldsymbol{\delta}}}$ and QAH are for comparison shown in Fig. \ref{fig:S_k}(b)-(f). The 20 Kekulé order parameters for $N=2$ fall into irreps $1,\bar{1}$ and $9,\bar{9}$ of the exact O(4). Here, $1$ and $\bar{1}$ are the spin-singlet bond density waves (BDW) whereas the other irreps contain spin triplet and charge BDW as well as SC Kekulés. It should be observed that the QSH and f-wave SC order parameters, which belong to the same irrep 9, show identical behavior. At the phase transition their structure factors change from a seemingly random distribution to a Gaussian-like shape in momentum space. However, in distinction to the order parameters that belong to the irrep 6, they fail to develop a sharp peak in the ordered phase. Finally, in stark contrast to both irreps 6 and 9, neither the Kekulé nor QAH order parameters' data shows any discernable sign at the phase transition.

	\section{Discussion}
	We now discuss our results in the context of the field-theoretic predictions and compare them with related work.
	
	The diverging structure factor of the Néel order parameter, given in Fig. \ref{fig:structurefactor} suggests that as argued in Refs. \cite{HanHerbut1,HanHerbut2} there is indeed a phase transition in the repulsive GN model. The observation that the critical interaction increases with the number of fermion components $N$ also agrees with the relevant Fierz identities and the mean-field analysis \cite{HerbutMandal}, which alone would predict a linear increase. The observed increase which appears faster than linear is presumably due to fluctuations, which should require a stronger than mean-field critical interaction to tame $\sim N^2$ of Goldstone bosons in the ordered state. We may also note that the critical attractive interaction for the transition into the QAH state should, in contrast, be essentially independent of $N$.  
	\begin{figure}
		\centering
		\includegraphics[width=0.99\columnwidth]{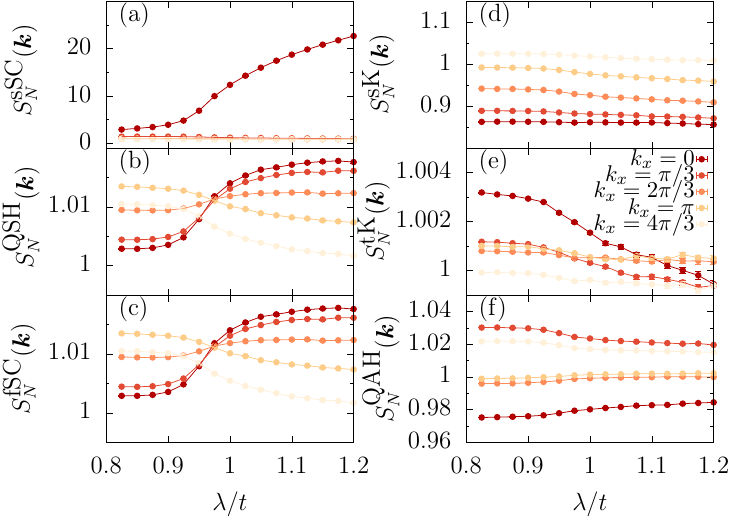}
		\caption{Normalized structure factor of  (a) sSC, (b) QSH, (c) fSC, (d) singlet Kekulé, (e) triplet Kekulé and (f) QAH at various wave vectors along the $k_x$-axis as a function of interaction $\lambda/t$ for $N=2$ and $L=12$ at half-filling.}
		\label{fig:S_k}
	\end{figure}

	When $N=2$ the transition appears to be almost continuous. However the poor quality of the data collapse and the suspicious value of the correlation length exponent needed to achieve it may be taken as evidence of a weakly discontinuous transition. This can be possibly understood as being due to a critical fixed point in the RG flow in the field theory with a small imaginary part \cite{HanHerbut2}. The fact that the transition becomes more discontinuous as $N$ increases can be understood from a renormalization group (RG) perspective \cite{HanHerbut2}. Although the transition is continuous at the mean-field level \cite{HerbutMandal}, the RG flow in the corresponding Gross-Neveu-Yukawa field theory for the tensor order parameter coupled to massless Dirac fermions shows a runaway into the region of couplings where the theory is unstable, which usually indicates a first-order transition. The RG ``time" it takes the flow to run away is inversely proportional to the size of the discontinuity \cite{HerbutBook}, and this was used to show in Ref. \cite{HanHerbut2} that for specific initial conditions the size increases with the number of fermionic components $N$. That the same behavior can be expected quite generally follows from the shape of the boundary between the regions of first- and second-order transitions in the $N-N_f$ plane \cite{HanHerbut2},  where $N_f$ is the additionally introduced number of fermion flavours, which in the present model is fixed at $N_f =1$: at any constant $N_f$ simply increasing $N$ takes one farther from the second-order and deeper into the first-order-transition territory.  It is also worth noticing that we find the transition to be discontinuous already at $N=2$, albeit only weakly so. This is in contrast to the result in another $\textrm{O}(4)$-symmetric model studied in Ref. \cite{Kaul}, but in agreement with the field-theoretic arguments  \cite{HerbutScherer, Uetrecht1, Uetrecht2}.  Both model Hamiltonians share the same symmetries and the numerical simulations point to quantum phase transitions between the very same phases. One hence expects similar criticality. A possible interpretation of this mismatch could lie in the weakly first order nature of the transition. In this case, the starting point in an RG  flow is determined by the  details of the interaction term and can lead to  very different flows, including quasi-criticality and strong first order transitions. An explicit example is discussed in \cite{Gotz}.

	%\fa{Is there not a contradiction here? After all, in Ribhu's work they have exactly the same symmetries obser a transition between the very same two phases. Would this not mean that Ribhu's model should also show a first order transition, but  that the model correspond  to a starting point in parameter space where the flow will be slower. After all for a weakly first order transition, with critical point in the complex plane, if you  start on the left of the critical point  then the flow may be very slow,  but if you start on the right then the flow  will be quick and you would see  a first order transition on small lattices.  They would  hence need bigger lattices to see this effect. } 
	
	The result that a finite chemical potential favors superconducting states is in accord with the mean-field analysis \cite{HerbutMandal}. In our exact O($2N$)-symmetric lattice model at half filling the insulating and the superconducting states that belong to the same irrep are exactly degenerate, and linear combinations of some of these are also possible. The ground state for general $N$ belongs to the representation given by the generators of O($2N$), and it is therefore $N(2N-1)$-dimensional. Since it is energetically preferable to gap out all the fermion components in the ordered state equally, the antisymmetric matrix in Eq. (\ref{eq:M2}) that corresponds to the ground state satisfies $A_{2N} ^2 = 1_{2N}$.  At half-filling therefore the exact O($2N$) symmetry becomes spontaneously broken down to U($N$). Chemical potential couples to the particle number, which is one of the generators of the O($2N$), and when finite manifestly reduces the symmetry to U($N$). The superconducting and the insulating states then no longer belong to the same irrep of the U($N$), and the mean-field calculation shows that the previously existing degeneracy is being broken in favor of superconductivity. If the O($2N$) symmetry were  already broken at half-filling in favor of insulators by some additional interactions, for example, a chemical potential above a finite critical value would be required to induce a flop into a superconducting state \cite{HerbutMandal}.  
	
	It is interesting to compare different order parameters in the lattice model at $N=2$. Consider the normalized structure factor in Fig. \ref{fig:S_k}. The sSC and Néel order parameter belong to the same irrep 6 of the exact lattice O($4$) symmetry. This is clearly observable in the data, as the spin and sSC structure factors are numerically equal. The same is true for the structure factors of fSC and QSH, which belong to the same irrep 9 of the O($4$), as seen in Fig. \ref{fig:S_k}(b) and Fig \ref{fig:S_k}(c). Although the structure factor for the irrep 9 develops a discernable peak at $\ve{k}=0$ at the transition, it is much broader than that of irrep 6 to which the order in the  ground state belongs. On the other hand, the normalized structure factor of the Kekulé order parameters, which belong to two (reducible) ten-dimensional representations, shown in Fig. \ref{fig:S_k}(d) remains completely flat. The different behavior of three different representations of the exact lattice O($4$) reflects the breaking of the low-energy O($8$) by the lattice terms: the irrep 35 of O($8$) 
	decomposes as 
	\begin{equation}
		35\rightarrow 6 + 9 + 10 + \bar{10},
	\end{equation}
	under the restriction to O($4$). In our model the irrep 6 order parameters develop long-range order in the ground state, while the order parameters in the irrep 9 seem to show some tendency to order, but ultimately do not; it would be interesting to see if in a different model it could be the other way around. 
	
Finally, in Fig. \ref{fig:S_k}(f) we see that the QAH order parameter, which is a singlet of both the exact O($4$) and the low-energy O($8$) remains 
short-ranged across the transition. This was to be expected, since the lattice model is constructed so that QAH order was to be present only when  the interaction is attractive. It would be illuminating, however, if one could circumvent the sign problem and confirm that the QAH state indeed develops for the negative interaction, as we argued in the paper.
% Alternative numerical methods may be helpful in this respect \cite{Kaul}. \fa{They have the very same limitations we have. So I would omit the sentence.}

In conclusion, our quantum Monte Carlo simulations confirm the existence of the O($4N$) symmetry-breaking phase transition on the repulsive side of the Gross-Neveu model, in broad 	agreement with field-theoretic predictions. The transition is weakly first-order for $N=2$ and becomes more strongly discontinuous with increasing $N$. The exact O($2N$) lattice symmetry produces a rich hierarchy of order parameters, with the irrep~6 (N\'{e}el/sSC/CDW) ordering while irrep~9 (QSH/fSC) remains disordered, and the Kekul\'{e} and QAH channels staying completely short-ranged.
	
	The QMC data together with the processed data used for generating the figures can be accessed online \cite{Osiander_Rein2026-data}
	
	\section{Acknowledgement} 
	
	 The authors would like to thank Emilie Huffman, Shailesh Chandrasekharan and Ribhu Kaul for discussions. FFA and IFH thank the Würzburg-Dresden Cluster of Excellence on Complexity and Topology in Quantum Matter ct.qmat (EXC 2147, project-id 390858490). GOR thanks the DFG for financial support under Grant No. AS 120/19-1 (Project No. 530989922). The authors gratefully acknowledge the Gauss Centre for Supercomputing e.V. (www.gauss-centre.eu) for funding this project by providing computing time on the GCS Supercomputer SUPERMUC-NG at Leibniz Supercomputing Centre (www.lrz.de). We also gratefully acknowledge the scientific support and HPC resources provided by the Erlangen National High Performance Computing Center (NHR@FAU) of the Friedrich-Alexander-Universität Erlangen-Nürnberg (FAU) under NHR project 80069 provided by federal and Bavarian state authorities. NHR@FAU hardware is partially funded by the German Research Foundation (DFG) through grant 440719683. IFH has also been supported by the NSERC of Canada.

	\appendix
	
	\section{The auxiliary field QMC method}
	This section's purpose is a brief, yet more detailed, overview over the finite temperature auxiliary field QMC method applied in this study. In order to make the lattice model defined in eq. \eqref{eq:Model} amenable to Markov Chain Monte Carlo (MCMC) sampling its partition fuction is first Trotterized to read
	\begin{equation}
		Z = \textrm{Tr}\left[\left(e^{-\Delta\tau \hat{H}_t}e^{-\Delta\tau \hat{H}_{\lambda}}\right)^{L_{\tau}}\right] + \mathcal{O}(\Delta\tau^2)
	\end{equation}
	where $\hat{H}_t$ is the kinetic (hopping) term, $\hat{H}_{\lambda}=\frac{2\lambda}{N}\sum_{n}\hat{V}_{n}^2$ with $\hat{V}_{n}=\sum_{\langle \langle \ve{i},\ve{j} \in n \rangle \rangle } \nu_{\ve{i}\ve{j}} c_{\ve{i}}^\dagger c_{\ve{j}}$ denotes the interaction and $L_{\tau}$ is the number of imaginary time steps to yield $L_{\tau}\Delta\tau = \beta$. The interaction part is decoupled by a Hubbard-Stratonovich transformation that for $\lambda>0$ reads
	\begin{equation}
		e^{-\frac{2\lambda\Delta\tau}{N}\hat{V}_n^2}
		\propto
		\int_{-\infty}^{\infty} d\chi_{n,\tau}\;
		e^{-\frac{N}{8\lambda}\Delta\tau\chi_{n,\tau}^2
			\;+\; \Delta\tau i\chi_{n,\tau}\hat{V}_n}
		\label{eq:HS}
	\end{equation}
	and introduces auxiliary fields $\chi_{n,\tau}$ that couple to $\hat{V}_n$ with a factor of $i$. This is due to the definition of the Gaussian integration
	$e^{a\hat{B}^2} \propto \int dx\,e^{-x^2/(4a) + x\hat{B}}$, only valid for $a>0$, that is applied with $\hat{B} = i\hat{V}_n$, $x=\Delta\tau\chi_{n,\tau}$ and $a=\frac{2\lambda\Delta\tau}{N}$. Note, that for $\lambda<0$ the second term in the above equation would simply be $\Delta\tau\chi_{n,\tau}\hat{V}_n$. What remains to arrive at the partition fuction in eq. \eqref{Eq:QMC} is to integrate out the fermions. This can be done by using the identity 
	\begin{equation}
		\textrm{Tr}\left[\prod_i e^{\boldsymbol{c}^\dagger M_i \boldsymbol{c}}\right]=\det\left[\mathbb{I}+\prod_ie^{M_i}\right]
	\end{equation}
	which applies for any hermitian or anti-hermitian matrix $M_i$ \cite{Assaad2008-QMC}. Doing this yields an action of the form
	\begin{equation}
		S = S_0 -  \ln\det \left[\mathbb{I}+\prod_{\tau}e^{-\Delta\tau M_t} \prod_{n}e^{\Delta\tau i \chi_{n,\tau}V_n}\right]
	\end{equation}
	where $M_t$ is the adjacency matrix connecting the sites of the lattice and $V_n$ is the matrix defining the interaction. The first term in the action $S_0$ is given by
	\begin{equation}
		S_0 = \frac{1}{8\lambda}\Delta\tau \sum_{n,\tau}\chi_{n,\tau}^2
	\end{equation}
	The partition fuction can thus be represented by 
	\begin{equation}
		Z = \sum_C e^{-N S(C)} + \mathcal{O}(\Delta\tau^2)
	\end{equation}
	where $C=\{\chi_{n,\tau}\}$ is a configuration of the Hubbard-Stratonovich fields.  As mentioned in the main text, the case $\lambda>0$ is sign-problem-free. For  $N=2$ this follows from the criterion of Ref. \cite{WuZhang}:  for  each  field configuration,  time  reversal symmetry as well as a charge  U(1)  symmetry  is present.  Hence,  the  eigenvalues of the fermion determinant enter in complex conjugate pairs.    It  then follows  that there is no 
	sign problem  for any even number of fermion flavors $N$.  
	As such, the determinant is real and non-negative for any configuration $C$, thus making $e^{-S(C)}$ a viable weight for MCMC.
	
	By sampling over the configuration space spanned by $C$ it is now possible to calculate observables using 
	\begin{equation}
		\langle \hat{O} \rangle = \frac{\textrm{Tr}\left(e^{-\beta \hat{H}} \hat{O}\right)}{\textrm{Tr}\left(e^{-\beta \hat{H}}\right)} = \sum_{C} \rho(C) \langle\langle O(C) \rangle\rangle_C
	\end{equation}
	with the distribution $\rho(C) = \frac{1}{Z} e^{-S(C)}$. In the practical application implemented in the ALF package \cite{ALF_v2.4}, the Hubbard-Stratonovich fields are discretized by applying Gauss-Hermite quadrature. Moreover, in order to guarantee hermiticity of the imaginary time propagation, we use a symmetric Trotter decomposition
	\begin{equation}
		e^{-\Delta\tau(\hat{H}_t+\hat{H}_\lambda)}
		=
		e^{-\frac{\Delta\tau}{2}\hat{H}_t}
		e^{-\Delta\tau\hat{H}_\lambda}
		e^{-\frac{\Delta\tau}{2}\hat{H}_t}
		+ \mathcal{O}(\Delta\tau^3),
	\end{equation}
	where the error of the order $\mathcal{O}(\Delta\tau^3)$ per time slice again yields an overall $\mathcal{O}(\Delta\tau^2)$ error in the partition fuction as $L_\tau\propto 1/\Delta\tau$. For further reading about the technical details of the technique the reader is referred to Refs. \cite{QMC-PhysRevB.40.506,QMC-PhysRevD.24.2278,Assaad2008-QMC} while the exact implementation of the algorithm is openly accessible via the ALF code \cite{ALF_v2.4}.
	
	\section{Superconducting order parameter}
	The on-site superconducting order parameter is defined as 
	\begin{equation}
		\hat{O}_{\boldsymbol{i}}^a=\frac{1}{2}c_{\boldsymbol{i},\sigma}^\dagger Y_{\sigma,\sigma'}^a c_{\boldsymbol{i},\sigma'}^\dagger+\textrm{h.c.}
	\end{equation} 
	where $Y^a$ is one of $a=N(N-1)/2$ possible matrices that can be chosen to define on-site pairing. In fact, only antisymmetric matrices $Y^a$ lead to non-zero values of $\hat{O}$ since
	\begin{eqnarray}
		&c_{\boldsymbol{i},\sigma}^\dagger Y_{\sigma,\sigma'}^a c_{\boldsymbol{i},\sigma'}^\dagger=(c_{\boldsymbol{i},\sigma}^\dagger Y_{\sigma,\sigma'}^a c_{\boldsymbol{i},\sigma'}^\dagger)^T = - c_{\boldsymbol{i},\sigma}^\dagger (Y_{\sigma,\sigma'}^a)^T c_{\boldsymbol{i},\sigma'}^\dagger \nonumber\\
		&	\Rightarrow Y^a = -(Y^a)^T.
	\end{eqnarray} 
	In the first step we use that this is just a bilinear and as such is the same as its transpose.
	$Y^a$ can be written as $Y^a=X^a \Sigma^a$
	with $\Sigma$ the unitary part of time reversal symmetry $T=\Sigma K$ ($K$ denoting the complex conjugation operator) and $\Sigma X^T \Sigma^{-1} = X$. The simplest possible choice would be $X=\mathbb{I}$ and $\mathbb{I}_{N-2}\otimes\sigma_{}^2$ thus restricting to an SU(2) spin degree of freedom. E.g. in the SU(4) case there are six possible matrices $X$ that satisfy this condition. These are $\mathbb{I}$ and the five Dirac matrices $\gamma_a$ with $a=1,\ldots,5$. Together they can be interpreted as the irreducible tensor representation of SU(4) but also form a 6-component vector that rotates under SO(6). By choosing $\Sigma = \mathbb{I}_2\otimes\sigma^2$ we can define the matrices $\gamma_a$ as $(\sigma^1\otimes\mathbb{I},\sigma^3\otimes\mathbb{I},\sigma^2\otimes\sigma^2,\sigma^2\otimes\sigma^3,\sigma^2\otimes\sigma^1)$ in order to satisfy the condition for $X$. In the SU(2) case there is only a single option for the matrix $Y$ which is $Y=\sigma^2$.\\ 

	In order to define the SC order parameter we may choose $\Sigma=\mathbb{I}_{N-2}\otimes\sigma^2$ and $X=\mathbb{I}$. Here, we show that regarding the computation of structure factors this is equivalent to choosing any other possible $Y$, i.e. the correlations function does not depend on $Y$. The correlation function reads
	\begin{eqnarray}
		\langle \hat{O}_{\boldsymbol{i}}^a\hat{O}_{\boldsymbol{j}}^b \rangle =  \frac{1}{4}\langle c_{\boldsymbol{i},\alpha}^\dagger Y_{\alpha,\beta}^a c_{\boldsymbol{i},\beta}^\dagger c_{\boldsymbol{j},\gamma} Y_{\gamma,\delta}^b c_{\boldsymbol{j},\delta}\rangle+\textrm{h.c.}
	\end{eqnarray}
	We can use the Wick decomposition in order to obtain
	\begin{align}
		\langle \hat{O}_{\boldsymbol{i}}^a\hat{O}_{\boldsymbol{j}}^b \rangle = & \frac{1}{4}\big(\langle c_{\boldsymbol{i},\alpha}^\dagger c_{\boldsymbol{j},\delta} \rangle \langle c_{\boldsymbol{i},\beta}^\dagger c_{\boldsymbol{j},\gamma} \rangle\nonumber\\ 
		&- 
		\langle c_{\boldsymbol{i},\alpha}^\dagger c_{\boldsymbol{j},\gamma} \rangle \langle c_{\boldsymbol{i},\beta}^\dagger c_{\boldsymbol{j},\delta} \rangle\big) Y_{\alpha,\beta}^aY_{\gamma,\delta}^b+\textrm{h.c.}
	\end{align}
	Due to SU($N$) spin symmetry the Green's function is diagonal and independent on the spin, i.e. $\langle c_{\boldsymbol{i},\sigma}^\dagger c_{\boldsymbol{j},\sigma'} \rangle=\langle c_{\boldsymbol{i}}^\dagger c_{\boldsymbol{j}} \rangle\delta_{\sigma,\sigma'}$. With this, the correlation value reads
	\begin{eqnarray}
		\langle \hat{O}_{\boldsymbol{i}}^a\hat{O}_{\boldsymbol{j}}^b \rangle = & \frac{1}{4}\langle c_{\boldsymbol{i}}^\dagger c_{\boldsymbol{j}} \rangle^2
		\big(Y_{\alpha,\beta}^aY_{\beta,\alpha}^b-Y_{\alpha,\beta}^aY_{\alpha,\beta}^b\big)+\textrm{h.c.}
	\end{eqnarray}
	Using the fact that $Y_{\alpha,\beta}^b=(Y_{\beta,\alpha}^b)^T=-Y_{\beta,\alpha}$ and choosing a basis where $\textrm{Tr}(Y^aY^b)=\delta^{a,b}$ we obtain 
	\begin{align}
		\langle \hat{O}_{\boldsymbol{i}}^a\hat{O}_{\boldsymbol{j}}^b \rangle = & \frac{1}{4}\langle c_{\boldsymbol{i}}^\dagger c_{\boldsymbol{j}} \rangle^2
		2 \, \textrm{Tr}\big(Y^aY^b\big)+\textrm{h.c.} \nonumber\\
		=&  \frac{N}{2} \delta^{a,b} \langle c_{\boldsymbol{i}}^\dagger c_{\boldsymbol{j}} \rangle^2+\textrm{h.c.}
	\end{align}
	This result shows that all the components $\hat{O}_{\boldsymbol{i}}^a$ lead to the same correlation value and enables us to compute it without specifying the matrix $Y^a$.

	\bibliography{./literature.bib}
	
\end{document}